\documentclass[preprint,authoryear,3p,times]{elsarticle}
\usepackage{graphicx}
\usepackage{psfrag}
\usepackage{subfigure}
\usepackage{color}
\usepackage{url}
\usepackage{stfloats}
\usepackage{amsmath}
\usepackage{array}

\usepackage{amscd}
\usepackage{amssymb}

\journal{Theoretical Biology}

\begin{document}

\begin{frontmatter}

\title{The $\sigma$ law of evolutionary dynamics in community-structured populations}

\author{$Changbing~Tang^{a},~Xiang~Li^{a}\footnote{\emph{Email address}:lix@fudan.edu.cn},~Lang~Cao^{b},~Jingyuan~Zhan^{a}$}

\address{$^{a}$Adaptive Networks and Control Lab, Department of Electronic Engineering Fudan University, Shanghai 200433, China\\
$^{b}$Department of Mathematical Engineering and Information Physics, University of Tokyo, Tokyo 153-8505, Japan}

\begin{abstract}
Evolutionary game dynamics in finite populations provides a new framework to understand the selection of traits with frequency-dependent fitness. Recently, a simple but fundamental law of evolutionary dynamics, which we call $\sigma$ law,  describes how to determine the selection between two competing strategies: in most evolutionary processes with two strategies, $A$ and $B$, strategy $A$ is favored over $B$ in weak selection if and only if $\sigma R + S > T + \sigma P$. This relationship holds for a wide variety of structured populations with mutation rate and weak selection under certain assumptions. In this paper, we propose a model of games based on a community-structured population and revisit this law under the Moran process. By calculating the average payoffs of $A$ and $B$ individuals with the method of effective sojourn time, we find that $\sigma$ features not only the structured population characteristics but also the reaction rate between individuals. That's to say, an interaction between two individuals are not uniform, and we can take $\sigma$ as a reaction rate between any two individuals with the same strategy. We verify this viewpoint by the modified replicator equation with non-uniform interaction rates in a simplified version of the prisoner's dilemma game (PDG).
\end{abstract}


\end{frontmatter}

\section{Introduction}
Evolutionary game theory was originally introduced as a tool for studying animal behavior \citep{Maynard (1973), Maynard (1982)} but has become a general approach that transcends almost every aspect of evolutionary biology \citep{Nowak (2004a)}, and provides a framework to understand the dynamics of frequency-dependent selection \citep{Hofbauer (1998), Gintis (2000), Cressman (2003), Nowak (2004a), Nowak (2006), Antal (2009a), Wu (2010), Gokhale (2011), Tarnita (2011)}.

The traditional approach of evolutionary game theory uses deterministic dynamics to describe infinitely large, well-mixed
populations \citep {Hofbauer (1988), Hofbauer (2003)}. To understand evolutionary game dynamics in finite-sized populations, a stochastic approach is developed \citep{Schaffer (1988), Fogel (1998), Ficici (2000), Alos-Ferrer (2003), Perc (2007), Traulsen (2007)}. A crucial quantity in the stochastic approach is the fixation probability of strategies which relies on the individual's reproduction ability \citep{Nowak (2004b), Taylor (2004), Imhof (2006), Nowak (2006), Traulsen (2006a), Lessard (2007), Bomze (2008)}. In the limit of weak selection, for a neutral mutant, probability $\rho_{A}$, that an individual using strategy $A$ will invade and take over the whole population of individuals using strategy $B$, is equal to the reciprocal of the population size, i.e., $1/N$. Therefore, the selection favors the fixation of invading strategy $A$ if $\rho_{A}>1/N$ under the weak selection.

Evolutionary game dynamics are also affected by the structure of population. Lots of early works focused on spatial games with regular lattices (Nowak and May, 1992; Nakamaru et al., 1997; Hauert and Doebeli, 2004; Szab\'{o} and Fath, 1998; Szab\'{o} et al., 2000), which recently, have been expanded to general structured populations, such as graphs \citep{Lieberman (2005), Ohtsuki (2006a), Ohtsuki (2006b), Ohtsuki (2007a), Taylor (2007), Cao (2008), Fu (2009), Li (2009), Perc (2010), Wang (2011)}, phenotype space \citep{Antal (2009b)} and set structured populations \citep{Tarnita (2009a)}. In a word, a population structure specifies who interacts with whom, and it greatly affects the outcome of an evolution. If the fitness of an individual is determined by its interactions with others, then we are in the world of evolutionary game theory \citep{Nathanson (2009)}.

Consider a game of two strategies, $A$ and $B$, in a population of fixed size $N$. Mutual cooperation is rewarded by $R$ for each player, whereas mutual defection pays each player a punishment $P$. If a cooperator plays against a defector, the former gets the sucker¡¯s payoff $S$ and the latter gets the temptation to defect $T$. Thus, the interactions are given by the payoff matrix
\begin{equation}
\begin{array}{ccc}
  ~& A & B \\
  A & R& S \\
  B & T& P \\
\end{array}
\end{equation}
Each individual obtains a payoff by interacting with others according to the population structure, and reproduces either genetically or culturally with a rate proportional to its payoff. Of course, reproduction is subject to mutation. Whenever an individual reproduces, the offspring adopts the parent's strategy with probability $1-\nu$ and a random strategy with probability $\nu$. To make the evolutionary dynamics of the process well approximated by an embedded Markov chain on the pure states \citep{Wu (2011)}, it's considered that mutation rate $\nu<(NlnN)^{-1}\ll 1$.

It is said that strategy $A$ is selected over strategy $B$ if it is more abundant in the stationary distribution of the mutation-selection process. \cite{Tarnita (2009b)} showed that with the weak selection, the condition that strategy $A$ is more abundant than strategy $B$ in the stationary distribution of the mutation-selection process is
\begin{equation}
\begin{array}{l}
\sigma R + S > T + \sigma P
\end{array}
\end{equation}
From this inequality, we know that the condition specifying which strategy is more
abundant is a linear inequality of the payoff values, $R$, $S$, $T$, $P$. It's pointed out that parameter $\sigma$, reflects how the structured populations influence the evolutionary dynamics \citep{Nowak (2010)}, and this inequality holds for a wide variety of population structures, including well mixed populations, graphs, phenotype space and set structured populations \citep{Nathanson (2009)}. Conveniently, we call this simple but fundamental law as the $\sigma$ law.

Recently, attention are focused on the population structures allowing selection at multiple levels \citep{Girvan (2002), Newman (2006), Traulsen (2005a), Traulsen (2006b), Traulsen (2008), Wang (2011), Hauert (2011), Wang (2012)}. In this case, a population is divided into groups (or communities), and the selection between individuals can be either in a community or among several communities. Following this line, we propose a model of games on a community-structured population based on the graph theory with a finite size. We verify that evolutionary games on the community-structured population obeys the $\sigma$ law from the viewpoint of interaction under the Moran process with sojourn time. Along the path of an invasion attempt, we find that the probability of an interaction between any two individuals with the same strategy is $\sigma$ times of that between different strategies. In other words, $\sigma$ represents the reaction rate between two individuals with the same strategy, and quantifies the degree that individuals using the same strategy are more ($\sigma>1$) or less ($\sigma<1$) likely to interact than individuals using different strategies.

\section{$\sigma$ law in community-structured populations}
\subsection{Model of community structures}
Consider a game played in a finite population of the fixed size $H$ distributed over $M$ communities (each community has the same size $N$, and any two of which have none common member) with strategies $A$ and $B$. Each individual obtains a payoff according to the payoff matrix $(1)$. At each time step, a single individual is selected proportional to its fitness for reproduction, and the offspring either replaces a randomly chosen individual in this community or moves to another community at any time step to replaces a randomly chosen individual. In \cite{Hauert (2011)}, the authors discussed the evolutionary dynamics in a deme structured population, and assumed that the migration rate is proportional to deme's fitness. In this paper, we suppose that there exists a graph in the structured population connecting communities, and the migration occurs between two communities if there is an edge between them (See Fig. 1).

Label all communities in the structured populations with $l=1, 2, \cdots, M$. The probability that a selected individual belongs to community $l$ migrates to community $m$ is given by $\lambda_{lm}$. Hence the migration process is determined by an $M\times M$ matrix, $\Lambda=[\lambda_{lm}]$, $\lambda_{ll}$ is the probability that an individual stays in community $l$. Note that $\lambda_{ll}+\sum_{m\in\Omega_{l}}^{M}\lambda_{lm}=1$, where $\Omega_{l}$ denotes the set of all the communities that individuals in community $l$ can migrate to. Thus $\Lambda$ is a stochastic matrix with an eigenvalue $1$, and no eigenvalue has an absolute value greater than $1$ \citep{Godsil (2001)}.

Assume that a mutant $A$-individual (individual using strategy $A$) is produced from a community $l$ with pure strategy $B$. This mutant replaces a randomly chosen individual in community $l$ with probability $\lambda_{ll}$, or migrates to community $m$ and replaces a randomly chosen individual in community $m$ with probability $\lambda_{lm}, m\in\Omega_{l}$. Since the graph connects all the communities as its nodes, there exists a path along which a mutant beginning in community $l, l\in\{1, 2, \cdots, M\}$ can be fixed in the whole population (note that the path is not unique). An example that the fixation process of a single mutant in the whole community-structured population is shown in Fig. 2, where the fixation of a mutant from community $1$ to take over the whole population corresponds to a path.

\subsection{$\sigma$ law in community-structured population with sojourn time}
Assume that a migration occurs very rarely and much less than mutation ($\lambda_{lm}\ll \nu<(NlnN)^{-1}\ll 1, m\in\Omega_{l})$. In the limit of $\lambda_{lm}\rightarrow 0$ and $\nu\rightarrow 0$, the population spends almost all the time in the homogeneous states with all individuals of type $A$ or type $B$. Since mutations occur much more frequently than migrations, a migrant has either invaded the whole community or been eliminated before the next migration takes place \citep{Traulsen (2006b)}. Due to this special structure, the interactions within a community is more frequent than those between communities. This allows us to use the method of sojourn time in the Moran process which was applied to explain the one-third law of evolutionary dynamics in \cite{Ohtsuki (2007b)}.

Suppose that a single individual from $l$-community is selected proportional to its fitness for reproduction. This newborn has to stay or migrate to another community and replaces a randomly selected individual, and the total number of the population keeps unchanged. Based on the migration matrix $\Lambda$ constructed previously, we know that the newborn stays in community $l$ with probability $\lambda_{ll}$, and migrate to community $m$ with probability $\lambda_{lm}$, where $\lambda_{ll}\gg \lambda_{lm}, m\neq l$.

First, we discuss the Moran process that happens in $l$-community with probability $\lambda_{ll}$. Among the total $N$ individuals in community $l$, $j$ of them follow strategy $A$, and $N-j$ follow strategy $B$. Introduce a parameter $\omega$ to measure the intensity of selection. The corresponding fitness of A and B individuals are given by
\begin{equation*}
\begin{array}{l}
f_{j}=1-\omega+\omega F_{j}\\
g_{j}=1-\omega+\omega G_{j}
\end{array}
\end{equation*}
where $F_{j}$ and $G_{j}$ are the expected payoffs for $A$ and $B$, respectively. In this paper, we consider the weak selection, i.e., $\omega\rightarrow 0$.
In the state space $\{0,\cdots,N\}$, there are two absorbing states, $j=0$ (which means $A$ has become extinct) and $j=N$ (which means $A$ has reached fixation). Then the transition probability $p_{j,j\pm1}$ is given by
\begin{equation}
\begin{array}{l}
p_{j,j+1}=(N-j)/N\cdot jf_{j}/[jf_{j}+(N-j)g_{j}]\\
p_{j,j-1}=j/N\cdot [(N-j)g_{j}]/[jf_{j}+(N-j)g_{j}]\\
p_{j,j}=1-p_{j,j+1}-p_{j,j-1}
\end{array}
\end{equation}
Consider a path along which a mutant invades a resident population from state $j$ $(1\leq j\leq N-1)$. Since the mutant eventually gets absorbed into either state $0$ or state $N$, we are interested in how much time is spent at state $j$ along the path. In \cite{Ohtsuki (2007b)}, the authors called it the sojourn time. Given all paths that start at state $i$, let $\bar{t}_{ij}$ be the mean sojourn time at state $j$ before the absorption into either state $0$ or state $N$, which is described as (see, for example, \cite{Ewens (2004)}).
\begin{equation}
\bar{t}_{ij}=\sum_{k=0}^{k^{l}}p_{ik}\bar{t}_{kj}+\delta_{ij},~~~\bar{t}_{0j}=\bar{t}_{Nj}=0
\end{equation}
In particular, if we start from a single mutant with no loss of generality, i.e.,
$i=1$, then Eq. (4) can be simplified as
\begin{equation}
\overline{t}_{1j}=N/j.
\end{equation}
This means that the stochastic process spends most of the time around the absorbing state, $j=0$.
Besides, the mean effective sojourn time $\overline{\tau}_{ij}$ at state $j$ for paths that start at state $i$ is given by $\overline{\tau}_{ij}=(p_{j,j+1}+p_{j,j-1})\overline{t_{ij}}$. Now, we only concern the mean effective sojourn time $\overline{\tau}_{1j}$ which starts from a single mutant under the neutral drift.
Since the fitness $f_{j}$ and $g_{j}$ are very close to $1$ under weak selection, with Eqs. (3) and (5), we obtain
\begin{equation}
\begin{array}{l}
\overline{\tau}_{1j}=(p_{j,j+1}+p_{j,j-1})\overline{t_{1j}}\\
=\{(N-j)/N\cdot jf_{j}/[jf_{j}+(N-j)g_{j}]+j/N\cdot [(N-j)g_{j}]/[jf_{j}+(N-j)g_{j}]\}N/j\\
\approx 2(N-j)/N
\end{array}
\end{equation}
We now can count the average number of games played by $A$-individual and $B$-individual along a path of invasion. At state $j$, an $A$-individual meets $(j-1)$ $A$-individuals and $(N-j)$ $B$-individuals, while a $B$-individual meets $j$ $A$-individuals and $(N-j-1)$ $B$-individuals. The average number of games played by $A$-individual and $B$-individual at state $j$ can be summarized as \citep{Ohtsuki (2007b)}
\begin{equation}
\left(
\begin{array}{cc}
  A\rightarrow A & A\rightarrow B \\
  B\rightarrow A & B\rightarrow B \\
\end{array}
\right)_{j}=
\left(
\begin{array}{cc}
  j-1 & N-j \\
  j & N-j-1 \\
\end{array}
\right)
\end{equation}
Therefore, the effective number of encounters are
\begin{equation*}
\sum_{j=1}^{N-1}\overline{\tau}_{1j}
\left(
\begin{array}{cc}
  j-1 & N-j \\
  j & N-j-1 \\
\end{array}
\right)=
\end{equation*}
\begin{equation*}2
\left(
\begin{array}{cc}
  \sum_{j=1}^{N-1}\frac{1}{N}(j-1)(N-j)&\sum_{j=1}^{N-1}\frac{1}{N}(N-j)(N-j) \\
  \sum_{j=1}^{N-1}\frac{1}{N}j(N-j)&\sum_{j=1}^{N-1}\frac{1}{N}(N-j-1)(N-j) \\
\end{array}
\right)
\end{equation*}
\begin{equation}
=2\cdot\left(
\begin{array}{cc}
  \sum_{j=1}^{N-1}\frac{1}{N}(j-1)N_{B}^{l}&\sum_{j=1}^{N-1}\frac{1}{N}(N-j)N_{B}^{l} \\
  \sum_{j=1}^{N-1}\frac{1}{N}(N-j)N_{A}^{l}&\sum_{j=1}^{N-1}\frac{1}{N}(j-1)N_{A}^{l} \\
\end{array}
\right)
\end{equation}
where $N_{A}^{l}$ is the number of $A$-individuals in community $l$ at state $j$, $N_{B}^{l}=N-N_{A}^{l}$ is the number of $B$-individuals in community $l$ at state $j$. Furthermore, denote $I_{XX}^{l}$ the total number of interactions that an $X$-individual interacts with other $X$-individuals, and $I_{XY}^{l} (X\neq Y)$ is the total number of interactions that an $X$-individual interacts with $Y$-individuals, where $X,Y\in\{A, B\}$. Symmetrically, we have the following equations $\langle I_{AA}^{l}N_{B}^{l}\rangle_{0}=\langle I_{BB}^{l}N_{A}^{l}\rangle_{0}$,  $\langle I_{AB}^{l}N_{B}^{l}\rangle_{0}=\langle I_{BA}^{l}N_{A}^{l}\rangle_{0}$ at neutrality \citep{Nathanson (2009)}. The notation $\langle \cdot \rangle_{0}$ denotes the quantity averaged over all states of the stochastic process under neutral drift, $w=0$.
Thus, Eq. (8) can be simplified as
\begin{equation*}
2\left(
\begin{array}{cc}
  \langle I_{AA}^{l}N_{B}^{l}\rangle_{0} & \langle I_{AB}^{l}N_{B}^{l}\rangle_{0} \\
  \langle I_{BA}^{l}N_{A}^{l}\rangle_{0} & \langle I_{BB}^{l}N_{A}^{l}\rangle_{0} \\
\end{array}
\right)
\end{equation*}
\begin{equation}
~~~\rightarrow~
\left(
\begin{array}{cc}
  \sigma^{l}  & 1 \\
  1        & \sigma^{l}\\
\end{array}
\right)
\end{equation}
where $\sigma^{l}=\langle I_{AA}^{l}N_{B}^{l}\rangle_{0}/\langle I_{AB}^{l}N_{B}^{l}\rangle_{0}$.
On the other hand, the newborn reproduced in community $l$ migrates to community $m$ with probability $\lambda_{lm}$. Suppose that in community $m$ there are $h$ $A$-individuals and $N-h$ $B$-individuals, where $N$ is the size of community $m$. Similarly, we can get the effective sojourn time $\overline{\tau}_{1h}\approx 2(N-h)/N$. And the effective number of encounters can be written as
\begin{equation}
\left(
\begin{array}{cc}
  \sigma^{m}  & 1 \\
  1        & \sigma^{m}\\
\end{array}
\right)
\end{equation}
where $\sigma^{m}=\langle I_{AA}^{m}N_{B}^{m}\rangle_{0}/\langle I_{AB}^{m}N_{B}^{m}\rangle_{0}$.

Since we only discuss the number of interactions along a path that the mutant invades the whole population, the total number of interactions is a linear composition of the number of interactions in every community. So the total effective number of interactions can be finally written as
\begin{equation}
\left(
\begin{array}{cc}
  \sigma  & 1 \\
  1        & \sigma\\
\end{array}
\right)
\end{equation}
where $\sigma=c_{l}\sigma^{l}+\sum_{m\in\Omega_{l}}^{M}c_{m}\sigma^{m}$. Here, the coefficient $c_{n} (n\in l\cup \Omega_{l})$,  reflects the strength of effect from effective number of interactions in community $n$ to total effective number of interactions.

Eq. (11) suggests that both types of individuals with the same strategy interact $\sigma$ times as often as they interact with the other individuals. Therefore the average payoffs of $A$ and $B$ individuals along an invasion-path in a community-structured population are $\sigma R+S$ and $T+\sigma P$, respectively. From this result we can get the $\sigma$ law under the weak selection in a community-structured population, i.e., the condition that strategy $A$ is more abundant than strategy $B$ in the stationary distribution is $\sigma R+S>T+\sigma P$.

\section{$\sigma$ law in the case of non-uniform interaction}
The $\sigma$ law provides a fundamental criterion to specify which strategy is more
abundant in a structured population. But, what's the $\sigma$? In \cite{Nowak (2010)}, the authors said that $\sigma$ is the `structure coefficient', and it reflects how the structured population influences the evolutionary dynamics. In this section, we point out that $\sigma$ also features the reaction rate between two individuals with the same strategy.

To verify this point, we will use the evolutionary game theory with non-uniform interaction rates discussed in \cite{Taylor (2006)}. They assumed that the probability of interaction between two individuals is dependent of their strategies, and described the interaction as
\begin{equation*}
\begin{array}{l}
A+A^{~\underrightarrow{~~r_{1}~~}~}AA\\
A+B^{~\underrightarrow{~~r_{2}~~}~}AB\\
B+B^{~\underrightarrow{~~r_{3}~~}~}BB
\end{array}
\end{equation*}
Let us consider a simplified version of the prisoner's dilemma game (PDG) with the payoff matrix as below
\begin{equation}
  \begin{array}{ccc}
     ~ & C & D \\
    C & b-c & -c \\
    D & b & 0 \\
  \end{array}
\end{equation}
If we discuss the game (12) in a population without community-structure, i.e. $M=1$, strategy $D$ dominates strategy $C$ with uniform interaction rates $r_{1} = r_{2} = r_{3}$. Eventually, the entire population consists of defectors. However, if individuals only interact with opponents of the same strategy, cooperators cannot be exploited by defectors. In this case, when $r_{2} = 0$ and $r_{1}, r_{3}> 0$, cooperation is the dominant strategy due to $b-c > 0$. Therefore, $r_{2}\neq 0$ means that cooperators and defectors do interact.

Without loss of generality, we assume that $r_{1}= r_{3}= r> 0$, $r_{2} = 1$ in a population with community-structure. Let $x^{l}$ and $y^{l}$ be the frequencies of individuals adopting strategy $C$ and $D$ in community $l$, respectively, and $x^{l} + y^{l} = 1$. The fitness of individuals are determined by the average payoffs over a large number of interactions. Therefore, the fitness of $A$ and $B$ individuals in community $l$ are
\begin{equation}
  \begin{array}{c}
f_{C}^{l}=[r(b-c)x^{l}-cy^{l}]/(r x^{l}+y^{l}), ~~~~f_{D}^{l}=bx^{l}/(x^{l}+ry^{l}),
  \end{array}
\end{equation}
In our model, an individual is randomly selected at each time step as a new vacancy, which is replaced either with the offspring of an individual from the same community or with an individual migrated from other communities. In community $l$, the probability of an $X$-individual ($X=C$, or $D$) filling a new vacancy due to local reproduction is proportional to the product of the number of $X$-individuals and their fitness, i.e., $N_{X}^{l}f_{X}^{l}$. And, the probability of an $X$-individual filling in a new vacancy due to global migration is proportional to the product of the averaged number of $X$-individuals and the migration rate $\lambda_{l}$, i.e., $\lambda_{l}\langle N_{X}\rangle$. Here, we suppose $\lambda_{lm}=\lambda_{l}=(1-\lambda_{ll})/N_{\Omega_{l}}, m\in \Omega_{l}$, where $N_{\Omega_{l}}$ is the number of set $\Omega_{l}$. Thus, we obtain the transition probabilities as
\begin{equation*}
  \begin{array}{l}
T_{C\rightarrow D}^{l}=(N_{C}^{l}/N)\cdot (N_{D}^{l}f_{D}^{l}+\lambda_{l}\langle N_{D}\rangle)/[\sum_{C,D}(N_{D}^{l}f_{D}^{l}+\lambda_{l}\langle N_{D}\rangle)]\\
~~~~~~~~~=(N_{C}^{l}/N)\cdot [\overline{f}^{l}/(\overline{f}^{l}+\lambda_{l})\cdot (N_{D}^{l}f_{D}^{l})/(N\overline{f}^{l})+\lambda_{l}/(\overline{f}^{l}+\lambda_{l})\cdot \langle N_{D}\rangle/N]
  \end{array}
\end{equation*}
\begin{equation}
  \begin{array}{l}
T_{D\rightarrow C}^{l}=(N_{C}^{l}/N)\cdot (N_{C}^{l}f_{C}^{l}+\lambda_{l}\langle N_{C}\rangle)/[\sum_{C,D}(N_{C}^{l}f_{C}^{l}+\lambda_{l}\langle N_{C}\rangle)]\\
~~~~~~~~~=(N_{D}^{l}/N)\cdot [\overline{f}^{l}/(\overline{f}^{l}+\lambda_{l})\cdot (N_{C}^{l}f_{C}^{l})/(N\overline{f}^{l})+\lambda_{l}/(\overline{f}^{l}+\lambda_{l})\cdot \langle N_{C}\rangle/N]
  \end{array}
\end{equation}
where $\overline{f}^{l}=f_{C}^{l}\cdot (N_{C}^{l}/N)+f_{D}^{l}\cdot (N_{D}^{l}/N)$, and $\langle N_{X}\rangle$=$\sum_{l=1}^{M}N_{X}^{l}/M $. From Eq. (14), we know that after a vacancy appears, either local reproduction occurs with probability $\overline{f}^{l}/(\overline{f}^{l}+\lambda_{l})$, or global migration occurs with probability $\lambda_{l}/(\overline{f}^{l}+\lambda_{l})$.

For sufficiently large but finite populations, Traulsen et al. have shown that the stochastic process can be well approximated by a set of stochastic differential equations combining deterministic dynamics and diffusion referred to as Langevin dynamics \citep{Traulsen (2005b), Traulsen (2006c)}. Hence from Eq. (14), we may derive
the Langevin equation to describe the evolutionary dynamics as
\begin{equation}
  \begin{array}{l}
\dot{x}^{l}=a^{l}(x^{l})+b^{l}(x^{l})\xi
  \end{array}
\end{equation}
where $a^{l}=T_{D\rightarrow C}^{l}-T_{C\rightarrow D}^{l}$, $b^{l}$ is the effective terms, and $\xi$ is the uncorrelated Gaussian noise. As $N\rightarrow \infty$, the diffusion term tends to zero and a deterministic equation is obtained
\begin{equation}
  \begin{array}{l}
\dot{x}^{l}=a^{l}(x^{l})=(N_{D}^{l}/N)\cdot [\overline{f}^{l}/(\overline{f}^{l}+\lambda_{l})\cdot (N_{C}^{l}f_{C}^{l})/(N\overline{f}^{l})+\lambda_{l}/(\overline{f}^{l}+\lambda_{l})\cdot \langle N_{C}\rangle/N]\\
~~~~~~-(N_{C}^{l}/N)\cdot [\overline{f}^{l}/(\overline{f}^{l}+\lambda_{l})\cdot (N_{D}^{l}f_{D}^{l})/(N\overline{f}^{l})+\lambda_{l}/(\overline{f}^{l}+\lambda_{l})\cdot \langle N_{D}\rangle/N]
  \end{array}
\end{equation}
Therefore, the equilibria of Eq. (16) which satisfy $T_{D\rightarrow C}^{l}=T_{C\rightarrow D}^{l}$ are
\begin{equation}
  \begin{array}{l}
[-(\alpha-2\beta)(x^{l})^{2}+(\alpha-2\beta)x^{l}+\beta]/[-(r-1)^{2}(x^{l})^{2}+(r-1)^{2}x^{l}+r]=\gamma/N
  \end{array}
\end{equation}
where $\alpha=r^{2}(b-c)-(b+c)$, $\beta=-rc$, and $\gamma=4(N_{C}^{l}-\overline{N}_{C})M\lambda_{l}$ with $\overline{N}_{C}=\sum_{i=1}^{M}N_{C}^{i}$.

Denote $\Delta N_{C}=N_{C}^{l}-\overline{N}_{C}$. When $\Delta N_{C}=0$, i.e., each community has the same numbers of $C$-individuals, Eq. (17) can be simplified as
\begin{equation}
  \begin{array}{l}
-(\alpha-2\beta)(x^{l})^{2}+(\alpha-2\beta)x^{l}+\beta=0
  \end{array}
\end{equation}
When $\alpha>-2\beta$, i.e., $r>(b+c)/(b-c)$, there are two interior equilibria.
Specifically, the stable interior equilibrium is
\begin{equation*}
  \begin{array}{c}
x^{l*}_{s}=1/2+\sqrt{\alpha^{2}-4\beta^{2}}/(2\alpha-4\beta)
  \end{array}
\end{equation*}
and the unstable interior equilibrium is
\begin{equation*}
  \begin{array}{c}
x^{l*}_{u}=1/2-\sqrt{\alpha^{2}-4\beta^{2}}/(2\alpha-4\beta)
  \end{array}
\end{equation*}
As $r$ increases, the two interior equilibria move symmetrically away from the bifurcation point $x^{l*}=1/2$, the interior stable equilibrium moves toward 1, and the unstable equilibrium moves toward 0. Therefore, the proportion of cooperators tends to increase monotonically with $r$ increases, and $(b+c)/(b-c)$ is the critical value which specifies whether cooperators are more abundant than defectors or not. That's to say, cooperators are more abundant than defectors when
\begin{equation}
r(b-c)-c>b
\end{equation}
This condition coincides with Eq. (2), for $R=b-c, S=-c, T=b, P=0$, and the interaction rate $r$ plays the same role as $\sigma$ in section 2.

Since the frequencies of individuals adopting strategy $C$ in each community become stable at the steady state of the evolutionary dynamics, $\Delta N_{C}\rightarrow 0$. Hence, when $\Delta N_{C}\neq 0$, we write $\Delta N_{C}=\Delta N_{C0}\cdot\varepsilon$  for some fixed $\Delta N_{C0}>0$, and Eq. (17) can be simplified as
\begin{equation}
  \begin{array}{l}
[\gamma(r-1)^{2}-N(\alpha-2\beta)](x^{l})^{2}-[\gamma(r-1)^{2}-N(\alpha-2\beta)]x^{l}+(N\beta-\gamma r)=0
  \end{array}
\end{equation}
When $N\alpha+2N\beta>\gamma(r+1)^{2}/N$, i.e., $r>(b+c+\gamma/N)/(b-c-\gamma/N)$, we obtain the stable interior equilibrium
\begin{equation*}
  \begin{array}{c}
x^{l*}_{s}=1/2+1/2\sqrt{[\gamma(r+1)^{2}-N(\alpha+2\beta)]/[\gamma(r-1)^{2}-N(\alpha-2\beta)]}
  \end{array}
\end{equation*}
and the unstable interior equilibrium is
\begin{equation*}
  \begin{array}{c}
x^{l*}_{u}=1/2-1/2\sqrt{[\gamma(r+1)^{2}-N(\alpha+2\beta)]/[\gamma(r-1)^{2}-N(\alpha-2\beta)]}
  \end{array}
\end{equation*}
This conclusion is similar to the case of $\Delta N_{C}= 0$, that is, the two interior equilibria move symmetrically away from the bifurcation point $x^{l*}=1/2$ and the proportion of cooperators tends to increase monotonically with $r$ increases. Thus, $(b+c+\gamma/N)/(b-c-\gamma/N)$ is the critical value which specifies whether cooperators are more abundant than defectors or not. With simplification, we have the condition that cooperators are more abundant than defectors
\begin{equation}
r(b-c)-c>b+(\gamma/N)(r+1)
\end{equation}
When $r>1$, a small $\varepsilon>0$ leads to $\gamma\rightarrow0$. Eq. (21) is discriminated from Eq. (19) only with the infinitesimal term, and the second term in the r.h.s of Eq. (21) is negligible in the meaning of mathematic. Therefore, Eq. (21) also coincides with Eq. (2), for $R=b-c, S=-c, T=b, P=0$.

To sum up, Eq. (19) and Eq. (21) reflect that the interaction rate $r$ plays the similar role as $\sigma$ in section 2, and we verify that $\sigma$ features the reaction rate between two individuals with the same strategy.

\section{Conclusion and Discussion}
In this paper, we have proposed a model of games on a community-structured population to understand the $\sigma$ law with the Moran process and the ergodic theory. By calculating the average payoffs of $A$ and $B$ individuals with the effective sojourn time method, we find that, $\sigma$ features not only the structured populations characteristics, but also the reaction rate between individuals. Parameter $\sigma$, reflects that individuals using the same strategy may interact with themselves $\sigma$ times than the interactions with individuals of the other strategy. That's to say, an interaction between two individuals are not uniform with the verification in a simplified PDG.

In practice, an overlap of communities usually exists in social networks for example, yet the case with overlap doesn't be discussed in this paper. Here we simplify an overlapped community-structured population to a none-overlap community-structured population. For a simple example without loss of generality, suppose that a game is played in a finite population with the fixed size $H$ distributed over $2$ communities, which have one common individual, as shown in Fig. 3(a). In this case, the overlap individual can be regarded as two parts, itself and a virtual reproduction which exist in the two corresponding separated communities, respectively. Therefore, the overlapped community-structured population with size $H$ becomes a none-overlap one with size $H+1$ (see Fig. 3(b)), which is discussed in our paper. Hence, we provide a simplified yet alternative approach to study the evolutionary games in a community-structured population with overlap.

\section*{Acknowledgment}
We were grateful to Lin Wang and Yu Wang for their helpful discussions and valuable suggestions, and the anonymous reviewers for their constructive comments to help improve this paper. This work is supported partly by NSFC (Grant No. 60874089), the National Key Basic Research and Development Program (Grant No. 2010CB731403), and the NCET program of China (Grant No. 09-0317).

\section*{References}

\bibliographystyle{model2-names}
\bibliography{<your-bib-database>}

\newpage
\begin{figure}[h]
\begin{center}
\begin{tabular}{c}
\includegraphics[height=4.5cm]{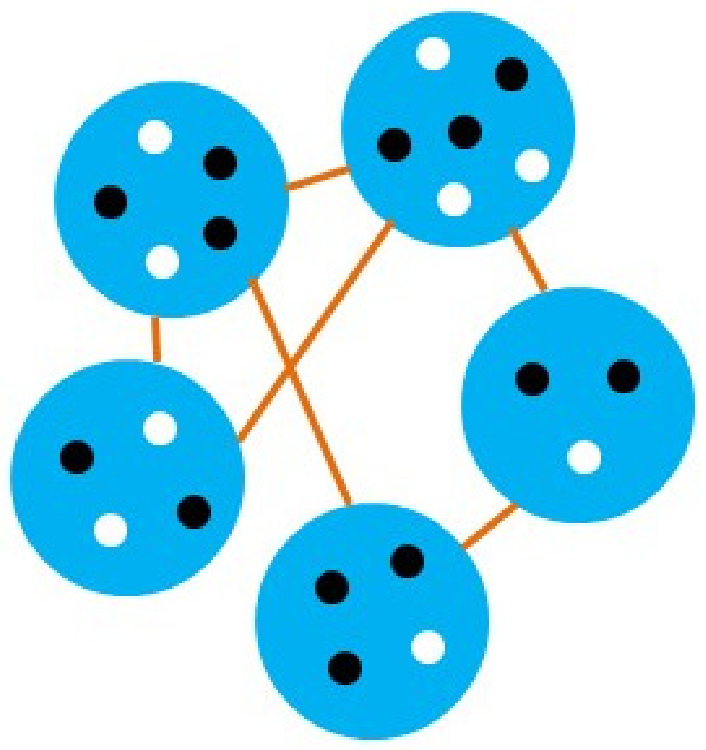}
\end{tabular}
\end{center}
\begin{bf}
Figure 1.
\end{bf}

An illustrative community-structured population with strategies $A$ and $B$. At each time step, a single individual from the entire population is selected proportional to its fitness for reproduction, and the offspring either stays in this community or migrates to another community. There exists a graph connecting all the communities in the structured population. The migration occurs between two communities if there is an edge connecting them both.
\end{figure}

\newpage
\begin{figure}[h]
\begin{center}
\begin{minipage}{7cm}
    \includegraphics[width=8.5cm]{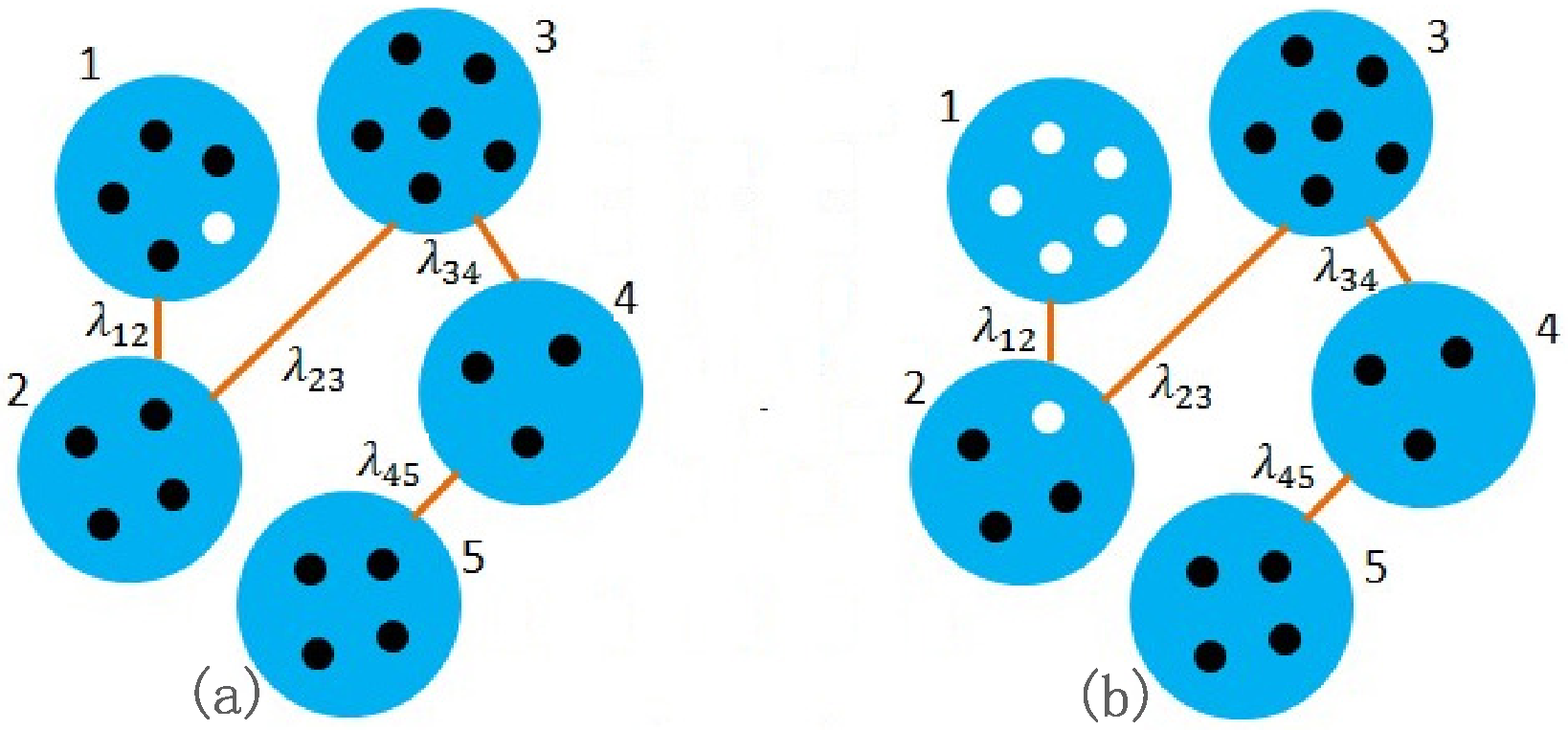}
  \end{minipage}\\
  \begin{minipage}{7cm}
    \includegraphics[width=8.5cm]{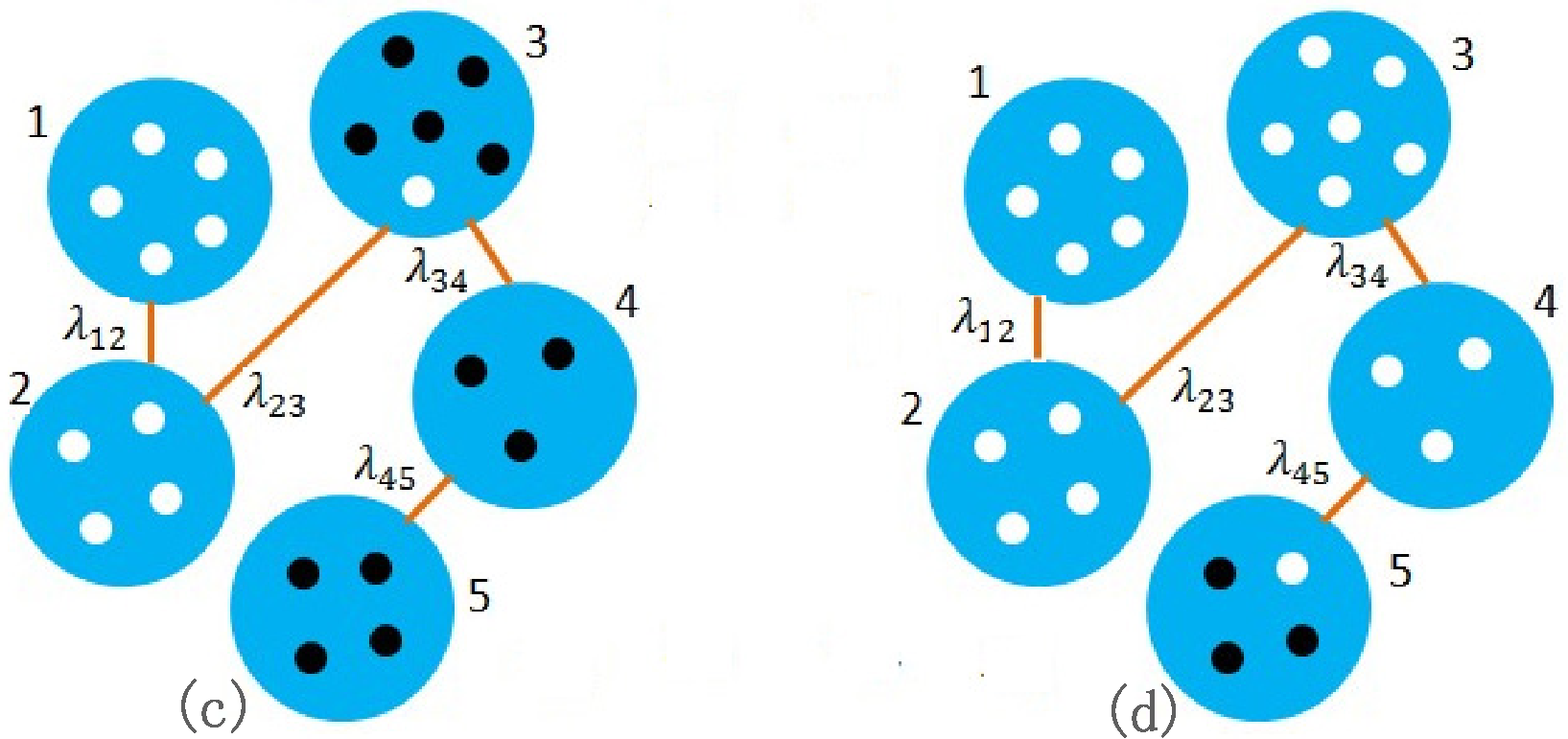}
  \end{minipage}
 \end{center}
\begin{bf}
Figure 2.
\end{bf}

An illustrative fixation process of a single mutant in the whole community-structured population. (a) A single mutant is produced in community $1$ (white node); (b) This mutant successfully takes over community $1$ with probability $\lambda_{11}$, and then migrates to community $2$ to replace a randomly chosen individual with probability $\lambda_{12}$; (c) The mutant takes over community $2$ with probability $\lambda_{22}$, and then migrates to community $3$ to replace a randomly chosen individual with probability $\lambda_{23}$; (d) The mutant takes over community $4$ with probability $\lambda_{44}$, and then migrates to community $5$ to replace a randomly chosen individual with probability $\lambda_{45}$.
\end{figure}

\newpage
\begin{figure}[h]
\begin{center}
\begin{tabular}{c}
\includegraphics[height=4.5cm]{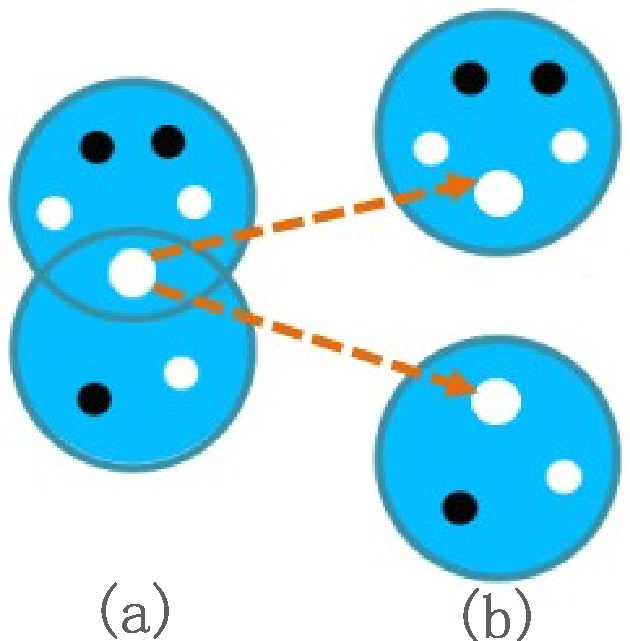}
\end{tabular}
\end{center}
\begin{bf}
Figure 3.
\end{bf}

(a) An overlapped community-structured population with one common individual; (b) A population with two none-common communities.
\end{figure}

\end{document}